\documentclass[conference]{IEEEtran}
\IEEEoverridecommandlockouts
\usepackage{cite}
\usepackage{amsmath,amssymb,amsfonts}
\usepackage{algorithmic}
\usepackage{graphicx}
\usepackage{textcomp}
\usepackage{xcolor}
\def\BibTeX{{\rm B\kern-.05em{\sc i\kern-.025em b}\kern-.08em
    T\kern-.1667em\lower.7ex\hbox{E}\kern-.125emX}}

\usepackage[labelformat=simple]{subcaption}

\IEEEspecialpapernotice{(Invited Paper)}
\newtheorem{remark}{Remark}

\begin{document}

\title{Coherent Compensation based ISAC Signal Processing for Long-range Sensing\\
}
\author{\IEEEauthorblockN{Lin Wang\textsuperscript{*}, Zhiqing Wei\textsuperscript{*}, Liyan Su\textsuperscript{\dag}, Zhiyong Feng\textsuperscript{*}, Huici Wu\textsuperscript{\ddag}  and Dongsheng Xue\textsuperscript{*}}\\
	\IEEEauthorblockA{\textsuperscript{*} Key
		Laboratory of Universal Wireless Communications, Ministry of Education\\
		Beijing University
		of Posts and Telecommunications, Beijing 100876, China\\
		\textsuperscript{\dag}Huawei Technologies Co., Ltd, Beijing 100085, China \\
		\textsuperscript{\ddag}National Engineering Lab for Mobile Network
		Technologies\\ Beijing University of Posts and Telecommunications, Beijing
		100876, China\\
		Email:\{wlwl, weizhiqing, fengzy, dailywu, xuedongsheng\}@bupt.edu.cn, suliyan1@huawei.com}}

\maketitle

\begin{abstract}
Integrated sensing and communication (ISAC) will greatly enhance the efficiency of physical resource utilization. The design of ISAC signal based on the orthogonal frequency division multiplex (OFDM) signal is the mainstream. However, when detecting the long-range target, the delay of echo signal exceeds CP duration, which will result in inter-symbol interference (ISI) and inter-carrier interference (ICI), limiting the sensing range. Facing the above problem, we propose to increase useful signal power through coherent compensation and improve the signal to interference plus noise power ratio (SINR) of each OFDM block. Compared with the traditional 2D-FFT algorithm, the improvement of SINR of range-doppler map (RDM) is verified by simulation, which will expand the sensing range. 
\end{abstract}


\section{Introduction}
6G wireless communication applications (autonomous driving, smart factory, etc) not only require communication but also require sensing the surrounding environment. It is necessary to integrate communication and sensing in the same system, which is called the integrated sensing and communication (ISAC). The ISAC will greatly improve the efficiency of physical resource utilization \cite{liu fan, andrew zhang, zhiyong feng}.

The ISAC technical routes include communication-centric design, radar-centric design and joint design. In which, the communication-centric ISAC refers to integrate sensing into communication system without influencing communication performances. The orthogonal frequency division multiplex (OFDM) widely used in the mobile communication system is the mainstream in ISAC signal design \cite{zhiqing wei}. While transmitting communication information, the range and velocity of the target can be estimated \cite{survey, braun, sturm}.  However, the sensing range of the OFDM signal is limited by the cyclic prefix (CP). For long-range target, there will be inter-symbol interference (ISI) and inter-carrier interference (ICI) in the echo signal\cite{no cp}. According to the CP duration defined in 3GPP TS 38.211, the maximum sensing range without ISI and ICI is only 88.5 meters, which greatly restricts the application of ISAC. Therefore, it is urgent to design an interference suppression method to expand the sensing range. 

Facing the above-mentioned challenges, Tigrek et al. \cite{no cp} removed the CP and transmitted the adjacent elementary OFDM symbols  continuously to reduce the Doppler ambiguity of long-range targets. Due to the absence of CP, OFDM signal is susceptible to multipath, which affects the communication. Hakobyan et al. \cite{repeated symbols,fd and delta_f} proposed repeated symbols OFDM (RS-OFDM), which only retains the CP of the first OFDM symbol and transmits the identical symbol consecutively. This scheme causes no ISI and ICI in the echo signal, and the maximum unambiguous speed is increased. But the repeated symbols limit the data rate. Tang et al. \cite{self interference} studied IEEE 802.11ad based sensing. When the delay of echo signal exceeds one OFDM symbol duration, the coherent superposition of adjacent symbols is used to accumulate energy. Then a small amount of interference-free OFDM symbols at the end of frame can be used for sensing. But it is necessary to insert additional pilots to assist sensing. Wu et al. \cite{vcp} proposed to re-segment the echo signal and transmitted signal into elementary sub-blocks in the time domain and adjacent sub-blocks can overlap. For echo signal, some samples called virtual cyclic prefix (VCP) after each sub-block are added onto the front of this sub-block to form a cyclic shift signal. The channel information matrix can be obtained by a point-wise division between fast fourier transformation (FFT) of re-segmented sub-block of echo signal and the transmitted signal. This scheme can flexibly re-segment the sub-blocks according to the sensing requirement. But these two schemes are not fully applicable to ISAC base station (BS): 1) echo delay exceeding one OFDM duration is not considered in ISAC BS; 2) the re-segmented sub-blocks will devastate the original OFDM symbol structure and increase the number of FFT. 

In this paper, ISAC BS is considered, which utilizes the 3GPP TS 38.211 based OFDM waveform. For the case where the echo delay exceeds CP duration but does not exceed one OFDM symbol duration, the ISI and ICI are modeled. Then, some samples after each elementary OFDM interval are added to the front through coherent compensation to increase power of the useful signal and improve the signal to interference plus noise power ratio (SINR) of each OFDM block. Since the data is processed within elementary OFDM symbol interval, the OFDM  structure is kept and the number of FFT will not be increased. The simulation results verify that the proposed coherent compensation scheme can improve the SINR of the range-doppler map (RDM) compared with the traditional 2D-FFT algorithm and VCP algorithm.

The rest of this paper is organized as follows. Section II introduces the OFDM signal model based on 3GPP TS 38.211 and models the ISI and ICI when the echo delay exceeds CP duration. Section III presents the proposed coherent compensation algorithm and derives the optimal compensation length. Section IV verifies the improvement of the proposed algorithm to the sensing performance through simulation. Section V concludes this paper.

\section{Signal Model}
\begin{figure}[htbp]
	\centerline{\includegraphics[scale=0.6]{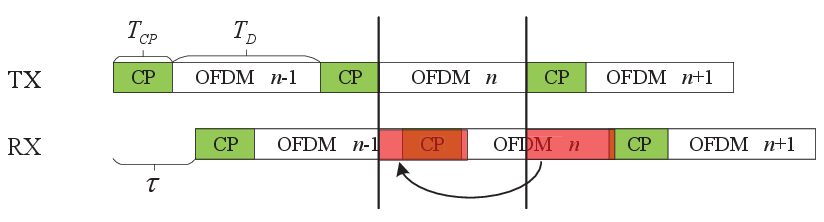}}
	\caption{OFDM echo signal and coherent compensation for target with echo delay exceeding CP duration. The samples (red part) after each elementary OFDM interval can be added to the front to improve the SINR.}
	\label{Fig. 1}
\end{figure}
The baseband signal of $M$ OFDM symbols (including CP) in the transmitter is
\begin{equation}
	x\left( t \right) = \frac{1}{{\sqrt {{N_c}} }}\sum\limits_{m = 0}^{M - 1} {\sum\limits_{k = 0}^{{N_c} - 1} {{S_m}\left( k \right){e^{j2\pi k\Delta f\left( {t - mT} \right)}}} u\left( {t - mT} \right)},
\end{equation}
where ${N_c}$ represents the number of subcarriers. ${S_m}\left( k \right)$ represents the normalized modulation symbol (e.g. QAM, PSK) transmitted on the $k$th subcarrier of the $m$th OFDM symbol. $\Delta f$ is the subcarrier spacing. $B = {N_c}\Delta f$ is the bandwidth. $T{\rm{ = }}{T_{CP}} + {T_D}$ is the duration of the entire OFDM symbol. ${T_{CP}}$ and ${T_D}$ are the duration of CP and elementary OFDM symbol, respectively. $u\left( t \right)$ is the pulse shaping function, generally using a rectangular function as follows.
\begin{equation}
	u\left( t \right) = \left\{ {\begin{array}{*{20}{c}}
			1&{ - {T_{CP}} \le t < {T_D}}\\
			0&{else}
	\end{array}}. \right.
	\label{rectangular function}
\end{equation}

Then the received echo baseband signal is
\begin{equation}
	y\left( t \right){\rm{ = }}\tilde \alpha x\left( {t - \tau } \right){e^{j2\pi {f_d}t}}{\rm{ + }}w\left( t \right).
\end{equation}
where $\tau$  and ${f_d}$  represent the time delay and Doppler shift caused by target.  $\tilde \alpha  = \alpha {e^{ - j2\pi {f_c}\tau }}$, where $\alpha $  represents the radar cross-section (RCS) of target, path loss and gain of antennas. $w\left( t \right)$ represents additive noise, which follows a stationary Gaussian random process. Since $x\left( t \right)$  is normalized, we can define the received SINR as  ${\gamma}_0 = \frac{{{{\left| {\tilde \alpha } \right|}^2}}}{{{\sigma ^2}}}$, where $\left| {\rm{*}} \right|$ represents the modulus operator and ${\sigma ^2}$ is the power of noise.

We consider that the radar receiver and transmitter are co-located and synchronized. For the case where the echo delay exceeds CP duration but is less than one OFDM symbol duration, i.e.  ${T_{CP}} < \tau  < T$, the $i$th sampling point with sampling interval ${T_s} = 1/B$  in the $n$th elementary OFDM symbol interval is \eqref{sampling in the nth elementary OFDM symbol interval}, 
\begin{figure*}[htbp]
	\begin{equation}
		{y_n}\left( i \right) 
		= y\left( {nT + i{T_s}} \right)
		= \frac{1}{{\sqrt {{N_c}} }}\sum\limits_{m = 0}^{M - 1} {\sum\limits_{k = 0}^{{N_c} - 1} {\tilde \alpha {S_m}\left( k \right){e^{j2\pi k\Delta f\left( {nT + i{T_s} - mT - \tau } \right)}}{e^{j2\pi {f_d}\left( {nT + i{T_s}} \right)}}} u\left( {nT + i{T_s} - mT - \tau } \right)} {\rm{ + }}{w_n}\left( i \right).
		\label{sampling in the nth elementary OFDM symbol interval}
	\end{equation}
\end{figure*}
where ${w_n}\left( i \right) 
= w\left( {nT + i{T_s}} \right)$, $i = 0,1, \cdots ,{N_c} - 1$. 	 From \eqref{rectangular function}, it can be seen that $ - {T_{CP}} \le nT + i{T_s} - mT - \tau  < {T_D}$  needs to be satisfied. Then we have $n + \frac{{i{T_s} - \tau  - {T_D}}}{T} < m \le n + \frac{{i{T_s} - \tau  + {T_{CP}}}}{T}$. Since  ${T_{CP}} < \tau  < T$, we can get  $m = n - 1,n$. So there will be ISI in $n$th elementary OFDM interval, as shown in Fig. \ref{Fig. 1}.  Doppler frequency shift will compromise the orthogonality between subcarriers. However, when the subcarrier spacing is considerably larger than Doppler frequency shift (usually satisfying $\Delta f > 10{f_d}$), the Doppler frequency shift on different subcarriers is considered to be equivalent, which ensures the orthogonality between subcarriers \cite{fd and delta_f}. Therefore, \eqref{sampling in the nth elementary OFDM symbol interval} can be transformed into \eqref{sampling v2 in the nth elementary OFDM symbol interval}.

%\begin{figure*}
	\begin{align}
			{y_n}&( i )
			= \frac{1}{{\sqrt {{N_c}} }}\sum\limits_{k = 0}^{{N_c} - 1} {\tilde \alpha {S_{n - 1}}\left( k \right){e^{j2\pi k\Delta f\left( {T + i{T_s} - \tau } \right)}}{e^{j2\pi {f_d}nT}}} \notag\\
			&\qquad\qquad\qquad\qquad\qquad\qquad\qquad \bullet u\left( {T + i{T_s} - \tau } \right)\notag\\
			&+ \frac{1}{{\sqrt {{N_c}} }}\sum\limits_{k = 0}^{{N_c} - 1} {\tilde \alpha {S_n}\left( k \right){e^{j2\pi k\Delta f\left( {i{T_s} - \tau } \right)}}{e^{j2\pi {f_d}nT}}u\left( {i{T_s} - \tau } \right)} \notag\\
			&+ {w_n}\left( i \right).
		\label{sampling v2 in the nth elementary OFDM symbol interval}
	\end{align}
%\end{figure*}

Let  ${N_s} = \lfloor {\tau /{T_s}}\rceil$ denote the sample offset corresponding to the delay of echo signal, where $\lfloor * \rceil$ denotes rounding to the nearest integer. ${N_{cp}} = \lfloor {{T_{CP}}/{T_s}} \rceil$  is the sample length of CP. It can be known from \eqref{rectangular function} that when $i = 0,1, \cdots ,{N_s} - {N_{cp}} - 1$,  $u\left( {T + i{T_s} - \tau } \right)$ is non-zero, and when  $i = {N_s} - {N_{cp}}, \cdots ,{N_c} - 1$, $u\left( {i{T_s} - \tau } \right)$ is non-zero. Performing FFT on \eqref{sampling v2 in the nth elementary OFDM symbol interval}, the frequency domain symbol on the $p$th subcarrier in the $n$th elementary OFDM symbol interval can be obtained as \eqref{frequency domain symbol}, where ${{W}_n}\left( p \right)$ is the FFT of ${w_n}\left( i \right)$.
\begin{figure*}
	\begin{align}
		{Y_n}\left( p \right) &= ( {1 - \frac{{{N_s} - {N_{cp}}}}{{{N_c}}}} )\tilde \alpha {S_n}\left( p \right){e^{\frac{{ - j2\pi p\tau }}{{N_c{T_s}}}}}{e^{j2\pi {f_d}nT}}+ \frac{1}{{{N_c}}}\sum\limits_{k = 0}^{{N_c} - 1} {[ {\tilde \alpha {S_{n - 1}}\left( k \right){e^{\frac{{j2\pi k\left( {{T_{CP}} - \tau } \right)}}{{{N_c}{T_s}}}}}( {\sum\limits_{i = 0}^{{N_s} - {N_{cp}} - 1} {{e^{\frac{{j2\pi \left( {k - p} \right)i}}{{{N_c}}}}}} } )} ]} {e^{j2\pi {f_d}nT}}\notag\\
		&+ \frac{1}{{{N_c}}}\sum\limits_{k = 0,k \ne p}^{{N_c} - 1} {[ {\tilde \alpha {S_n}\left( k \right){e^{\frac{{ - j2\pi k\tau }}{{{N_c}{T_s}}}}}( {\sum\limits_{i = {N_s} - {N_{cp}}}^{{N_c} - 1} {{e^{\frac{{j2\pi \left( {k - p} \right)i}}{{{N_c}}}}}} } )} ]} {e^{j2\pi {f_d}nT}}+ {W_n}\left( p \right).
		\label{frequency domain symbol}
	\end{align}
{\noindent} \rule[-10pt]{18cm}{0.05em}
\end{figure*}
It can be seen from \eqref{frequency domain symbol} that the average power of the useful signal, ISI and ICI are
\begin{equation}
	{P_u} = {( {1 - \frac{{{N_s} - {N_{cp}}}}{{{N_c}}}})^2}{\left| {\tilde \alpha } \right|^2},
\end{equation}
\begin{equation}
	{P_{ISI}} = \frac{{{{\left| {\tilde \alpha } \right|}^2}}}{{N_c^{}}}\left( {{N_s} - {N_{cp}}} \right),
	\label{ISI before compensation}
\end{equation}
\begin{equation}
	{P_{ICI}} = \frac{{{{\left| {\tilde \alpha } \right|}^2}}}{{N_c^2}}\left( {{N_c} - {N_s} + {N_{cp}}} \right)\left( {{N_s} - {N_{cp}}} \right).
\end{equation}
The SINR of the received frequency domain signal is
\begin{equation}
	\varUpsilon_f = \frac{{{{( {1 - \frac{{{N_s} - {N_{cp}}}}{{{N_c}}}} )^2}}}}{{2\frac{{{N_s} - {N_{cp}}}}{{N_c^{}}} - {{( {\frac{{{N_s} - {N_{cp}}}}{{N_c^{}}}} )^2}} + \frac{{{1}}}{{{{\gamma}_0}}}}}.
	\label{frequency domain SINR before coherent compensation}
\end{equation}
Therefore, as the echo delay (or $N_s$) increases, the frequency domain SINR decreases when the $\gamma_0$  is fixed.

\section{Coherent Compensation Based ISAC Signal Processing}
When the echo delay exceeds  CP duration, the SINR of the received frequency domain symbol decreases due to the existence of ISI and ICI. In which, the power of useful signal attenuates with the increase of the delay, while the ISI sees the same tendency along with delay.  When the delay is $\tau  = {T_D}/2 + {T_{CP}}$, orthogonality between subcarriers is most severely damaged, which leads to the most serious ICI. The orthogonality between subcarriers is destroyed due to not receiving a complete OFDM symbol. In order to improve the SINR, the $N'$  samples after the elementary OFDM symbol interval can be added to the front of this interval, as shown in Fig. \ref{Fig. 1}. The $l$th sample after $n$th elementary OFDM symbol is \eqref{compensation sampling},
\begin{figure*}
	\begin{flalign}
		\hspace{-6em}
		\label{compensation sampling} y\left( {nT + \left( {l + {N_c}} \right){T_s}} \right)
		&\approx \frac{1}{{\sqrt {{N_c}} }}\sum\limits_{k = 0}^{{N_c} - 1} {\tilde \alpha {S_n}\left( k \right){e^{j2\pi k\Delta f\left( {l{T_s} - \tau } \right)}}{e^{j2\pi {f_d}nT}}u\left( {l{T_s} + {N_c}{T_s} - \tau } \right)} \\
		&+ \frac{1}{{\sqrt {{N_c}} }}\sum\limits_{k = 0}^{{N_c} - 1} {\tilde \alpha {S_{n + 1}}\left( k \right){e^{j2\pi k\Delta f\left( {l{T_s} - {T_{CP}} - \tau } \right)}}{e^{j2\pi {f_d}nT}}u\left( {l{T_s} + {N_c}{T_s} - T - \tau } \right)}+ {w_n}\left( l \right).\notag
	\end{flalign}
\end{figure*}
where  $l = 0,1, \cdots ,N' - 1$. 

It can be seen from \eqref{compensation sampling} that when  $N' \le {N_s}$, the samples come from the $n$th OFDM symbol. Therefore, the sample compensation can realize the coherent combination of the energy and increase the SINR. When  $N' > {N_s}$, the interference from the ($n+1$)th OFDM symbol will be introduced. The $i$th sample in the $n$th elementary OFDM symbol interval after coherent compensation when $N' \le {N_s}$ is \eqref{sampling after compensation}.
\begin{figure*}
	\begin{align}
		 {{\tilde y}_n}&( i )	= \frac{1}{{\sqrt {{N_c}} }}\sum\limits_{k = 0}^{{N_c} - 1} {\tilde \alpha {S_{n - 1}}\left( k \right){e^{j2\pi k\Delta f\left( {T + i{T_s} - \tau } \right)}}{e^{j2\pi {f_d}nT}}} u\left( {T + i{T_s} - \tau } \right) + {{\tilde w}_n}\left( i \right)  \label{sampling after compensation} \\  
		&+ \frac{1}{{\sqrt {{N_c}} }}\sum\limits_{k = 0}^{{N_c} - 1} {\tilde \alpha {S_n}\left( k \right){e^{j2\pi k\Delta f\left( {i{T_s} - \tau } \right)}}{e^{j2\pi {f_d}nT}}} u( {\frac{{Ti}}{{N' - 1}} - {T_{CP}}})+ \frac{1}{{\sqrt {{N_c}} }}\sum\limits_{k = 0}^{{N_c} - 1} {\tilde \alpha {S_n}\left( k \right){e^{j2\pi k\Delta f\left( {i{T_s} - \tau } \right)}}{e^{j2\pi {f_d}nT}}u\left( {i{T_s} - \tau } \right)} \notag
	\end{align}
\end{figure*}
The noise of different samples is i.i.d, so the sample compensation is an incoherent combination for noise. The noise distribution after compensation is
\begin{equation}
 {\tilde w_n}\left( i \right){\thicksim}\left\{ {\begin{array}{*{20}{l}}
 		{N( {0,2{\sigma ^2}} )}&{0 \le i \le N' - 1}\\
 		{N( {0,{\sigma ^2}} )}&{N' \le i \le {N_c} - 1}
 \end{array}} \right.
	\label{noise distribution after compensation}
\end{equation}

Performing FFT on \eqref{sampling after compensation}, the frequency domain symbol after coherent compensation on the $p$th subcarrier in the $n$th elementary OFDM symbol interval is \eqref{frequency symbol after compensation}, where ${{\tilde W}_n}\left( p \right)$ is the FFT of ${\tilde w_n}\left( i \right)$.
\begin{figure*}
	\begin{align}
		{{\tilde Y}_n}\left( p \right)
		&= ( {1 - \frac{{{N_s} - {N_{cp}}}}{{{N_c}}} + \frac{{N'}}{{{N_c}}}})\tilde \alpha {S_n}\left( p \right){e^{\frac{{ - j2\pi p\tau }}{{N_c{T_s}}}}}{e^{j2\pi {f_d}nT}}+ \frac{1}{{{N_c}}}\sum\limits_{k = 0}^{{N_c} - 1} {[ {\tilde \alpha {S_{n - 1}}\left( k \right){e^{\frac{{j2\pi k\left( {{T_{CP}} - \tau } \right)}}{{{N_c}{T_s}}}}}( {\sum\limits_{i = 0}^{{N_s} - {N_{cp}} - 1} {{e^{\frac{{j2\pi \left( {k - p} \right)i}}{{{N_c}}}}}} } )} ]} {e^{j2\pi {f_d}nT}}\notag\\
		&+ \frac{1}{{{N_c}}}\sum\limits_{k = 0,k \ne p}^{{N_c} - 1} {[ {\tilde \alpha {S_n}\left( k \right){e^{\frac{{ - j2\pi k\tau }}{{{N_c}{T_s}}}}}( {\sum\limits_{i = 0}^{N' - 1} {{e^{\frac{{j2\pi \left( {k - p} \right)i}}{{{N_c}}}}}}  + \sum\limits_{i = {N_s} - {N_{cp}}}^{{N_c} - 1} {{e^{\frac{{j2\pi \left( {k - p} \right)i}}{{{N_c}}}}}} } )} ]} {e^{j2\pi {f_d}nT}}+ {{\tilde W}_n}\left( p \right)
		\label{frequency symbol after compensation}
	\end{align}
\end{figure*}
From \eqref{frequency symbol after compensation}, we can get the average power of the useful signal, ISI, ICI, and noise after coherent compensation as follows.
\begin{equation}
	{P_u} = {( {1 - \frac{{{N_s} - {N_{cp}}}}{{{N_c}}} + \frac{{N'}}{{{N_c}}}} )^2}{\left| {\tilde \alpha } \right|^2},
	\label{useful power after compensation}
\end{equation}
\begin{equation}
	{P_{ISI}} = \frac{{{{\left| {\tilde \alpha } \right|}^2}}}{{N_c^{}}}\left( {{N_s} - {N_{cp}}} \right),
	\label{ISI after compensation}
\end{equation}
\begin{equation}
	{P_{ICI}} = \frac{{{{\left| {\tilde \alpha } \right|}^2}}}{{N_c^2}}\left| {{N_s} - {N_{cp}} - N'} \right|\left( {{N_c} - \left| {{N_s} - {N_{cp}} - N'} \right|} \right),
	\label{ICI after compensation}
\end{equation}
\begin{equation}
	{P_w} = \frac{{{N_c} + N'}}{{{N_c}}}{\sigma ^2}.
\end{equation}
Then the SINR of the received frequency domain symbol after coherent compensation is \eqref{frequency domain SINR after compensation}.
\begin{figure*}
	\begin{equation}
		\varUpsilon_f = \frac{{{{(1 - \frac{{{N_s} - {N_{cp}}}}{{{N_c}}} + \frac{{N'}}{{{N_c}}})}^2}}}{{\frac{{{N_s} - {N_{cp}}}}{{{N_c}}} + \frac{1}{{N_c^2}}\left| {{N_s} - {N_{cp}} - N'} \right|\left( {{N_c} - \left| {{N_s} - {N_{cp}} - N'} \right|} \right) + \frac{{{N_c} + N'}}{{{N_c}}}\frac{1}{{{\gamma _0}}}}}
		\label{frequency domain SINR after compensation}
	\end{equation}
{\noindent} \rule[-10pt]{18cm}{0.05em}
\end{figure*}

Comparing \eqref{ISI before compensation} and \eqref{ISI after compensation}, it can be seen that when  $N' \le {N_s}$, coherent compensation does not affect the power of ISI. This is because the ISI comes from previous OFDM symbol, which is not affected during coherent compensation. It can be seen from \eqref{ICI after compensation} that the ICI disappears when  $N'{\rm{ = }}{N_s} - {N_{cp}}$. Because a complete OFDM symbol is formed.  When $N'$  continues to increase, the extra samples cannot maintain orthogonality in the FFT interval, causing ICI to appear again. But the extra samples can still increase the power of useful signal through coherent compensation, which can be seen from \eqref{useful power after compensation}. When  $N' > {N_s}$, since the adjacent OFDM symbols are independent, the compensated samples of the next OFDM symbol will not increase the power of useful signal, but will increase the power of ISI. Therefore, the SINR will be lower compared with the case when  $N' = {N_s}$. Next, we try to analyze the optimal sample compensation length.

Let  ${N_e} = \frac{{{N_s} - {N_{cp}}}}{{{N_c}}}$,  ${N_a} = \frac{{N'}}{{{N_c}}}$, then \eqref{frequency domain SINR after compensation} can be expressed as
\begin{equation}
	\varUpsilon_f = \frac{{{{\left( {1 - {N_e} + {N_a}} \right)}^2}}}{{{N_e} + \left| {{N_e} - {N_a}} \right|\left( {1 - \left| {{N_e} - {N_a}} \right|} \right) + \left( {1 + {N_a}} \right)\frac{1}{{\gamma{_0}}}}}.
	\label{frequency domain SINR after compensation simple}
\end{equation}

\subsection{${N_a} \le {N_e}$, i.e. $N' \le {N_s} - {N_{cp}}$}\label{AA}
When  ${N_a} \le {N_e}$ (i.e.  $N' \le {N_s} - {N_{cp}}$), \eqref{frequency domain SINR after compensation simple} can be expressed as
\begin{equation}
	\varUpsilon_f = \frac{{{{\left( {1 - {N_e} + {N_a}} \right)}^2}}}{{\left( {2{N_e} - {N_a}} \right) - {{\left( {{N_e} - {N_a}} \right)}^2} + \left( {1 + {N_a}} \right)\frac{1}{{\gamma{_0}}}}}.
	\label{frequency domain SINR after compensation Na<Ne}
\end{equation}
Deriving \eqref{frequency domain SINR after compensation Na<Ne} with respect to  ${N_a}$, we can get  $\frac{\partial }{{\partial {N_a}}}\varUpsilon_f > 0$. That is to say, when the number of compensated samples is in the range of $[0, N_s-N_{cp}]$, $\varUpsilon_f$ reaches maximum by compensating $N_s-N_{cp}$ samples.

\subsection{${N_a} > {N_e}$, i.e. ${N_s} - {N_{cp}} < N' \le {N_s}$}
When  ${N_a} > {N_e}$ (i.e. ${N_s} - {N_{cp}} < N' \le {N_s}$), \eqref{frequency domain SINR after compensation simple} can be expressed as
\begin{equation}
	\varUpsilon_f = \frac{{{{\left( {1 - {N_e} + {N_a}} \right)}^2}}}{{{N_a} - {{\left( {{N_a} - {N_e}} \right)}^2} + \left( {1 + {N_a}} \right)\frac{1}{{\gamma{_0}}}}}.
	\label{frequency domain SINR after compensation Na>Ne}
\end{equation}
Deriving \eqref{frequency domain SINR after compensation Na>Ne} with respect to  ${N_a}$, we can get
\begin{equation}
	\frac{\partial }{{\partial {N_a}}}{\varUpsilon _f} = \frac{{\left( {1 - {N_e} + {N_a}} \right)g\left( {{N_a}} \right)}}{{{{\left[ {{N_a} - {{\left( {{N_a} - {N_e}} \right)}^2} + \left( {1 + {N_a}} \right)\frac{1}{{{\gamma _0}}}} \right]}^2}}}
\end{equation}
where
\begin{equation}
	g\left( {{N_a}} \right) = {N_a}( {3 + \frac{1}{{\gamma{_0}}}} ) + \left( {1 + {N_e}} \right)\frac{1}{{\gamma{_0}}} - \left( {1 + {N_e}} \right).
\end{equation}

Since $g\left( {{N_a}} \right)$  is an increasing function with respect to  ${N_a}$, $\varUpsilon_f$ is maximal when $N' = {N_s} - {N_{cp}}$  or  $N' = {N_s}$ and can be expressed as in \eqref{sinr Na=Ne} and \eqref{sinr Na=Ns}, respectively.
\begin{equation}
	\varUpsilon_f\left( {{N_a}{\rm{ = }}{N_e}} \right) = \frac{{{N_c}}}{{\left( {{N_s} - {N_{cp}}} \right) + \left( {{N_c} + {N_s} - {N_{cp}}} \right)\frac{1}{{\gamma{_0}}}}}.
	\label{sinr Na=Ne}
\end{equation}
\begin{equation}
	\varUpsilon_f( {{N_a}{\rm{ = }}\frac{{{N_s}}}{{{N_c}}}}) = \frac{{{{\left( {{N_c} + {N_{cp}}} \right)}^2}}}{{{N_c}{N_s} - N_{cp}^2 + {N_c}\left( {{N_c} + {N_s}} \right)\frac{1}{{\gamma{_0}}}}}.
	\label{sinr Na=Ns}
\end{equation}
\begin{itemize}
\item $\varUpsilon_f( {{N_a}{\rm{ = }}\frac{{{N_s}}}{{{N_c}}}} ) > \varUpsilon_f\left( {{N_a}{\rm{ = }}{N_e}} \right)$

When  $d > d_0$, where
\begin{equation}
	d_0=\frac{{T_D^2 + {T_D}{T_{CP}} + T_{CP}^2}}{{2{T_D} + {T_{CP}}}}\frac{{{c_0}}}{2},
	\label{d0}
\end{equation}
$\varUpsilon_f( {{N_a}{\rm{ = }}\frac{{{N_s}}}{{{N_c}}}} ) > \varUpsilon_f\left( {{N_a}{\rm{ = }}{N_e}} \right)$  is always satisfied. For  $d < d_0$, the received SINR needs to satisfy	$\gamma{_0} < \gamma $, where
\begin{equation}
	\gamma {\rm{ = }}\frac{{\left( {{N_c} + {N_{cp}}} \right)\left( {{N_c}+{N_s} - {N_{cp}}} \right)  + {N_c}\left( {{N_s} - {N_{cp}}} \right)}}{{{{\left( {{N_c} + {N_{cp}}} \right)}^2} - {N_s}\left( {{N_c} + {N_{cp}}} \right) - {N_c}\left( {{N_s} + {N_{cp}}} \right)}}.
\end{equation}

\item {$\varUpsilon_f( {{N_{a}}{\rm{ = }}\frac{{{N_s}}}{{{N_c}}}} ) < \varUpsilon_f\left( {{N_{a}}{\rm{ = }}{N_{e}}} \right)$}

When  $d < d_0$, the received SINR needs to satisfy $\gamma{_0} > \gamma$.
\end{itemize}

\begin{remark}
	\begin{itemize}
		\item[(a)]  When the compensated sample  $0 < N' \le {N_s} - {N_{cp}}$, the SINR increases with increasing $N'$.
		\item[(b)]  When the target distance  $d > d_0$, regardless of the received SINR $\gamma{_0}$, the SINR after coherent compensation is the largest when  $N' = {N_s}$.
		\item[(c)] 	When the target distance  $d <d_0$, the SINR after $N' = {N_s}$ coherent compensation is the largest when  $\gamma{_0} < \gamma $. If $\gamma{_0} > \gamma $, the SINR is the largest when compensating $N' = {N_s} - {N_{cp}}$  samples.
	\end{itemize} 
\end{remark}

After point-wise dividing ${S_n}\left( p \right)$  in \eqref{frequency symbol after compensation}, the channel information can be obtained. The index corresponding to the peak in the RDM can be obtained in \eqref{range index} and \eqref{velocity index} through the 2D-FFT algorithm, as shown in Fig. 2.
\begin{equation}
	\hat k{\rm{ = }}\lfloor {\frac{\tau }{{{T_s}}}} \rceil.
	\label{range index}
\end{equation}
\begin{equation}
	\hat l{\rm{ = }}\lfloor {{f_d}MT} \rceil.
	\label{velocity index}
\end{equation}
Then, we can get the range and speed of the target as follows.
\begin{equation}
	r = \frac{{{c_0}\hat k}}{{2B}}.
\end{equation}
\begin{equation}
	v = \frac{{{c_0}\hat l}}{{2{f_c}M({N_c} + {N_{cp}}){T_s}}}.
\end{equation}

\begin{figure}[htbp]
	\centerline{\includegraphics[scale=0.5]{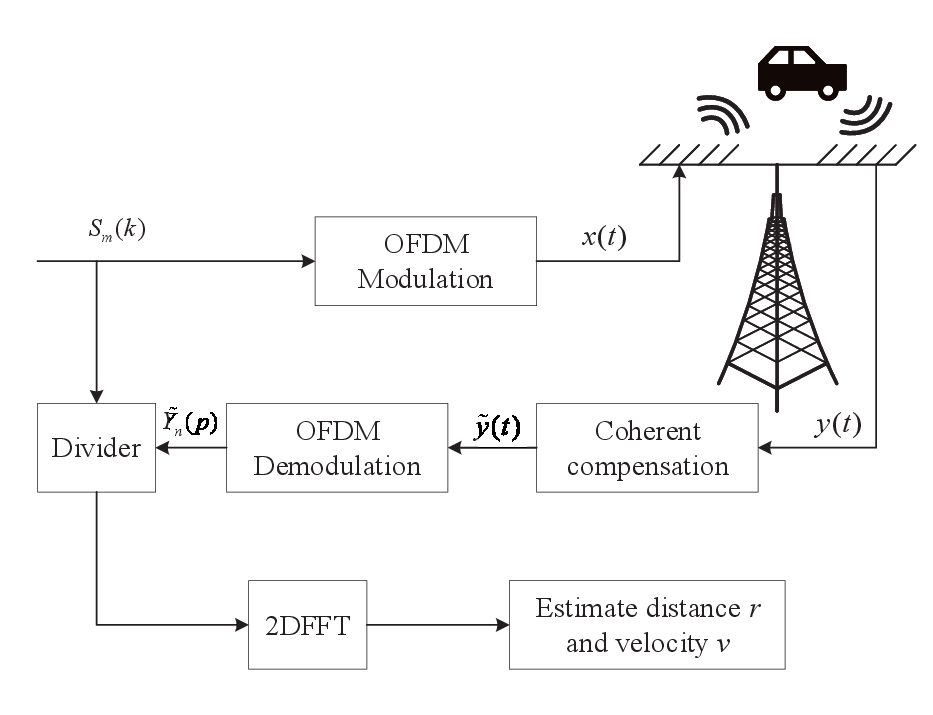}}
	\caption{Sensing signal processing for target with echo delay exceeding CP duration.}
	\label{Fig. 2}
	
\end{figure}

\section{Simulation	Results}
\begin{table}[h]
	\caption{\label{sys_para}Simulation parameters \cite{3GPP 38.211}}
	\begin{center}
		\begin{tabular}{l l}
			\hline
			\hline
			
			{Parameter} & {Value}\\
			
			\hline
			
			Carrier frequency ${f_c}$ & $28$ GHz\\
			Subcarrier spacing $\Delta f$ & $120$ kHz\\
			Number of subcarriers $N_c$ & 4096 \\
			Number of OFDM symbols $M$ & 160\\
			Elementary OFDM symbol duration ${T_D}$ & $8.33$ $\mu$s\\
			CP duration ${T_{CP}}$ & $0.59$ $\mu$s\\
			Entire OFDM symbol duration $T$ & $8.92$ $\mu$s\\
			Modulation & 16QAM\\
			\hline
			\hline
		\end{tabular}
	\end{center}
	\label{table_1}
\end{table}
The simulation parameters are shown in TABLE \ref{table_1}, which come from 3GPP TS 38.211. The speed of the target is not our main concern, so it is set to $20$ m/s in this paper. For BS-based communication, there will be no ISI and ICI if the maximum delay extension doesn't exceed the CP duration. But for BS-based sensing, it is required that the echo delay doesn't exceed the CP duration. Due to the two-way characteristic, the coverage of sensing is much smaller than that of communication. So there will be a higher  probability of the occurrence of ICI and ISI in BS-based sensing. It can be seen from TABLE \ref{table_1} that under this configuration, the sensing range corresponding to the CP duration is only $88.5$ m. If there is no special statement, the received  SINR of echo signal of the $500$ m target is set to $2$ dB, and the received SINR of echo signal of target at other distance is calculated by the LOS path loss model in urban macro scenario as follows \cite{3GPP 38.901}.
\begin{equation}
	{\rm{Pathloss}}\left[ {dB} \right] = 28.0 + 22{\log _{10}}(d) + 20{\log _{10}}({f_c}).
\end{equation}
where $f_c$  is in GHz and the distance  $d$ between the transmitting and receiving antennas is in meters. In the following simulations, we use the SINR of RDM as a metric of performance.
\begin{figure}[htb]
	\centering
	\begin{subfigure}[b]{0.23\textwidth}
		\centering
		\includegraphics[scale=0.3]{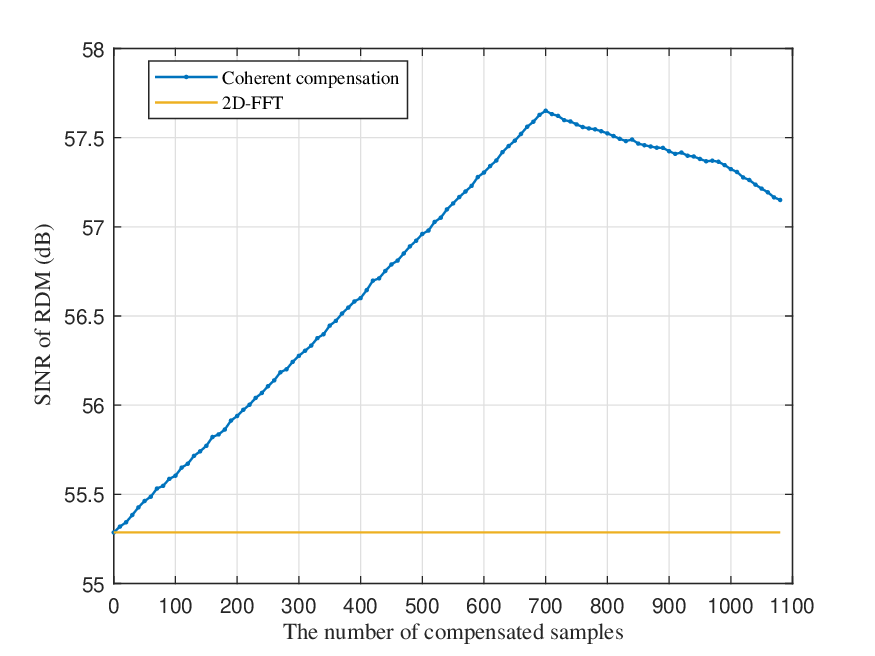}
		\caption{$300$ m target}
		\label{Fig. 3(a)}
	\end{subfigure}%
	\hfill
	\begin{subfigure}[b]{0.23\textwidth}
		\centering
		\includegraphics[scale=0.3]{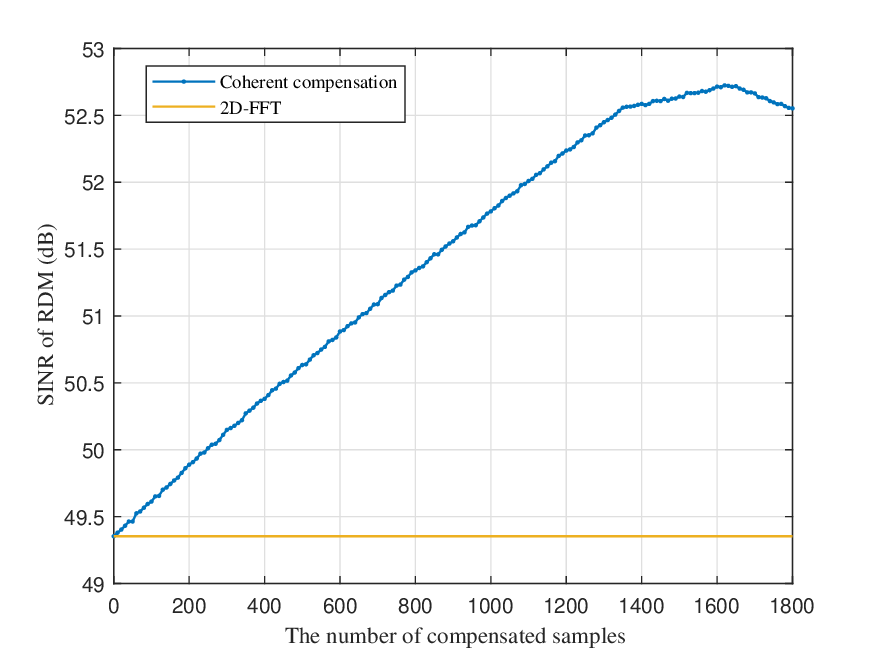}
		\caption{$500$ m target}
		\label{Fig. 3(b)}
	\end{subfigure}%
	% \vspace{-0.3cm}
	\caption{The SINR of RDM of target at $300$ m and $500$ m vs. The number of compensated samples.}
	\label{Fig. 3}
\end{figure}

\begin{figure}[htb]
	\centering
	\includegraphics[scale=0.5]{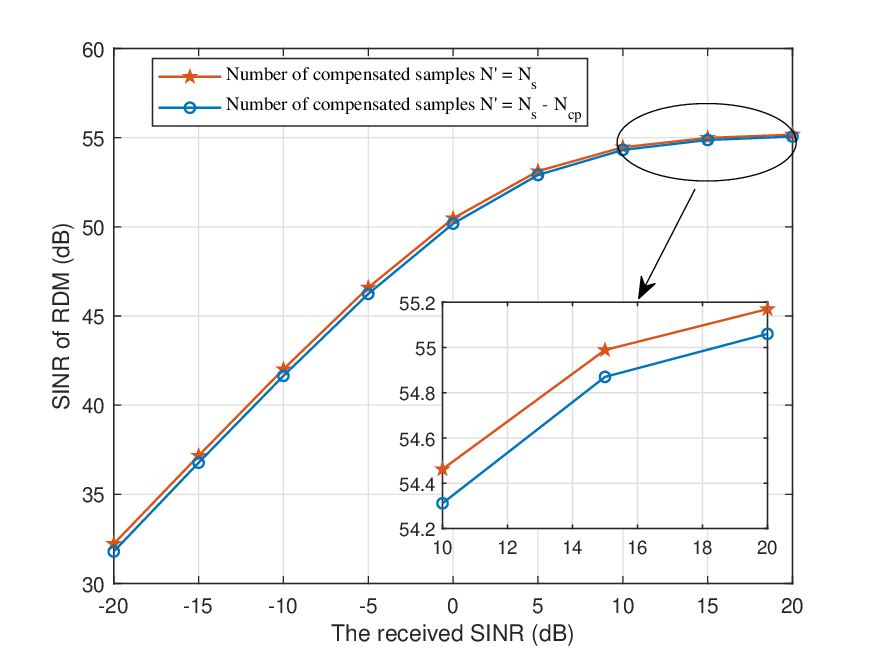}
	\caption{{Comparison of two compensated sample lengths for $800$ m target at different received SINR. }}
	\label{Fig. 4}
\end{figure}

Fig. \ref{Fig. 3} shows the SINR of RDM versus the number of compensated samples when the target is located at $300$ m with received SINR $\gamma_0=11.76$ dB and $500$ m with received SINR $\gamma_0=2$ dB. It can be seen from Fig. \ref{Fig. 3(a)} and Fig. \ref{Fig. 3(b)} that the SINR of RDM increases as the number of compensated samples increases when $N' \le {N_s} - {N_{cp}}$, which verifies the \emph{Remark} 1(a). When the number of compensated samples is  $N'{\rm{ = }}{N_s} - {N_{cp}}$, SINR of $300$ m target's RDM is maximized. But for $500$ m target, the best length of compensated samples is $N'{\rm{ = }}{N_s}$, verifying the \emph{Remark} 1(c). Fig. \ref{Fig. 4} depicts the comparison of two compensation lengths for $800$ m target at different received SINR. From Fig. \ref{Fig. 4} we can see, no matter what the received SINR is, the SINR of $800$ m target's RDM is maximized when  $N'{\rm{ = }}{N_s}$, which verifies the \emph{Remark} 1(b).

\begin{figure}[htb]
	\centering
	\includegraphics[scale=0.45]{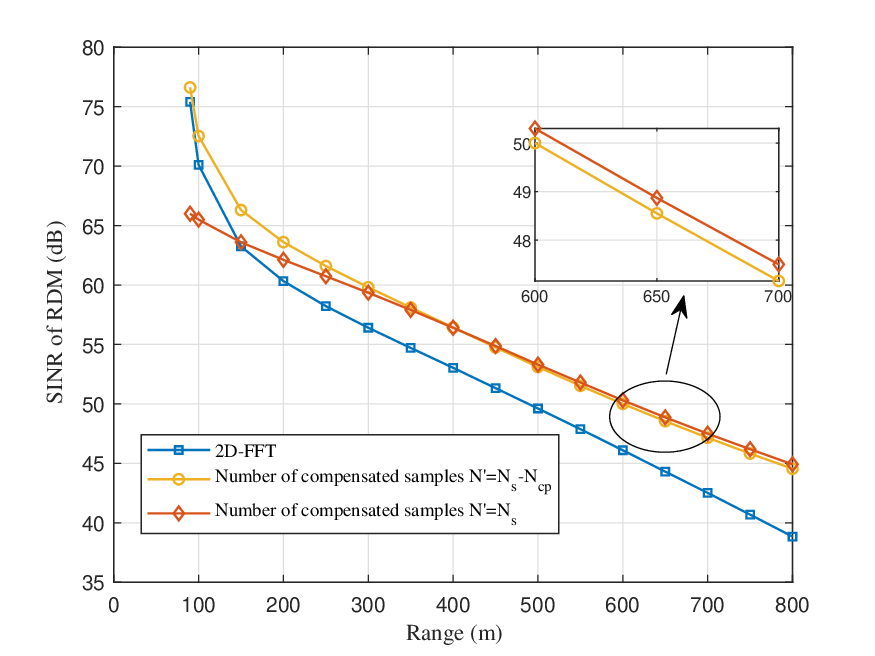}
	\caption{{SINR of targert's RDM at different ranges. }}
	\label{Fig. 5}
\end{figure}

Fig. \ref{Fig. 5} depicts the SINR of RDM obtained by 2D-FFT, coherent compensation for $N' = {N_s} - {N_{cp}}$  and $N' = {N_s}$  when the target is located at different ranges (echo delay exceeds CP duration). It can be seen from Fig. \ref{Fig. 5} that the SINR of RDM obtained by compensating $N' = {N_s} - {N_{cp}}$ samples is better than that of traditional 2D-FFT algorithm. Because there is no ICI at this case, and the power of useful signal (i.e. peak of RDM) is increased through coherent compensation. When the target is relatively close, compensating $N' = {N_s} $ samples will reduce the SINR of RDM. Because the received SINR of the echo is high, and the sample compensation will greatly enlarge the noise of RDM, which lowers the SINR of RDM. When the target is far away, the SINR of RDM obtained by compensating $N' = {N_s}$ samples is better than that by compensating $N' = {N_s} - {N_{cp}}$ samples.

\begin{figure}[htb]
	\centering
	\includegraphics[scale=0.45]{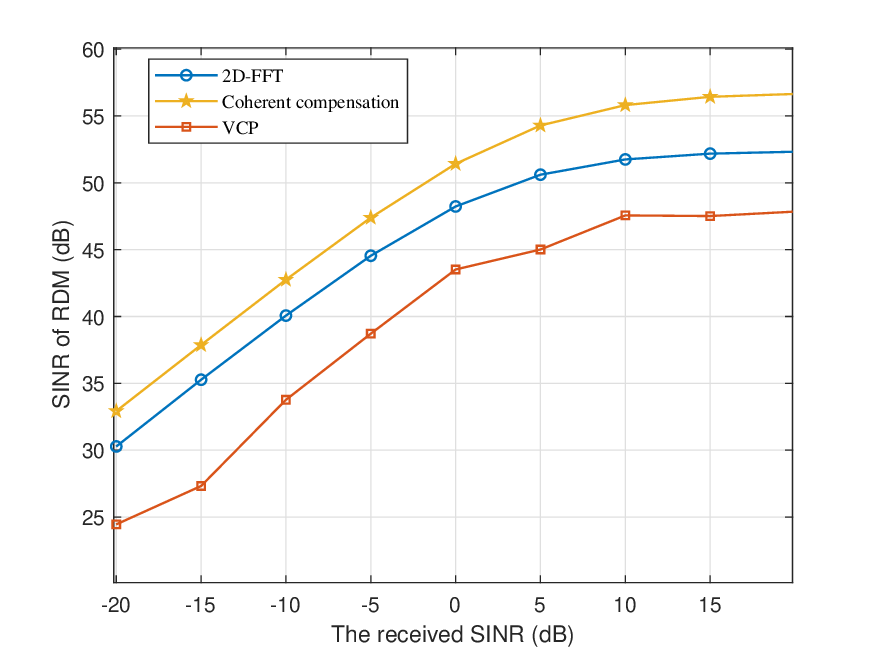}
	\caption{{SINR of RDM of target at $500$ m  under different sensing algorithms }}
	\label{Fig. 6}
\end{figure}
Fig. \ref{Fig. 6} demonstrates the SINR of  RDM of target at $500$ m obtained by 2D-FFT, VCP, and coherent compensation  under different received SINR. The VCP scheme re-segments each sub-block into $12$ OFDM symbols, and $4$ OFDM symbols overlap between adjacent sub-blocks. The length of VCP is the sample offset corresponding to the echo delay. It can be seen from the figure that compared with the traditional 2D-FFT algorithm, the coherent compensation scheme improves the SINR of RDM by $2\sim4$ dB. However, it's worth mentioning that the VCP scheme is less efficacious when constellation symbols are transmitted in the frequency domain. This is because the re-segmentation of the echo signal destroys the original OFDM structure, causing the symbols recovered by FFT to roughly follow the Gaussian distribution. Therefore, the probability of recovered frequency domain symbols with smaller amplitudes is higher. The noise will be amplified in the process of point-wise division, thus worsening the SINR of RDM.

\begin{figure}[htb]
	\centering
	\includegraphics[scale=0.45]{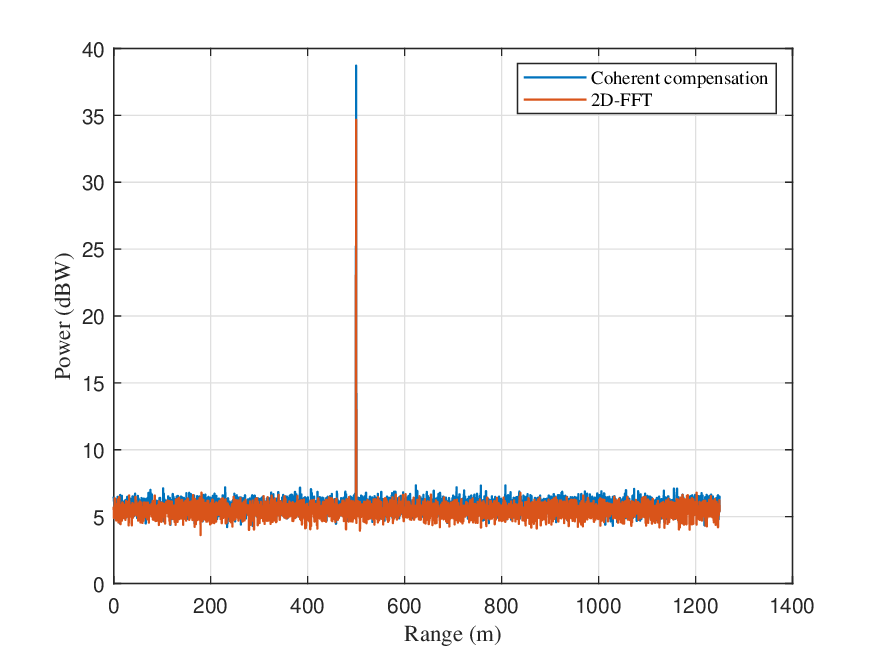}
	\caption{ {Range profile of target at 500m. }}
	\label{Fig. 7}
\end{figure}
Fig. \ref{Fig. 7} plots the range profile obtained by coherent compensation and 2D-FFT when the target is located at $500$ m. Compared with 2D-FFT scheme, coherent compensation scheme increases the peak power, which eventually enlarges the SINR in the range profile, albeit with the noise floor slightly raised simultaneously.

\section{Conclusion}
This paper studies ISAC BS, which utilizes the 3GPP TS 38.211 based OFDM waveform. When the echo delay exceeds CP duration, there will be ISI and ICI. The impaired SINR will limit sensing range. We propose to add some samples after each elementary OFDM interval to the front through coherent compensation to improve the SINR of each OFDM block. The simulation results show that the proposed algorithm improves the SINR of RDM compared with traditional 2D-FFT algorithm, which will expand the sensing range.

\section*{Acknowledgment}

This work was supported in part by the National Natural Science Foundation of China (NSFC) under Grant U21B2014, 62271081, and 92267202, and in part by the National Key Research and Development Program of China under Grant 2020YFA0711302.


\begin{thebibliography}{00}
	
\bibitem{liu fan} F. Liu et al., “Integrated sensing and communications: Toward dual-functional wireless networks for 6G and beyond,” 
{\em IEEE J. Sel. Areas Commun.}, vol. 40, no. 6, pp. 1728-1767, Jun. 2022.

\bibitem{andrew zhang} J. A. Zhang et al., “An overview of signal processing techniques for joint communication and radar sensing,” {\em IEEE J. Sel. Top. Sign. Proces.}, vol. 15, no. 6, pp. 1295-1315, Nov. 2021.

\bibitem{zhiyong feng} Z. Feng, Z. Fang, Z. Wei, X. Chen, Z. Quan and D. Ji, “Joint radar and communication: A survey,” {\em China Communications}, vol. 17, no. 1, pp. 1-27, Jan. 2020.

\bibitem{zhiqing wei} Z. Wei et al., “Integrated sensing and communication signals toward 5G-A and 6G: A Survey,” {\em IEEE Internet Things J.}, vol. 10, no. 13, pp. 11068-11092, 1 Jul., 2023.

\bibitem{survey} W. Zhou, R. Zhang, G. Chen and W. Wu, “Integrated sensing and communication waveform design: A survey,” {\em 	IEEE open J. Commun. Soc.}, vol. 3, pp. 1930-1949, Oct. 2022.

\bibitem{braun} M. Braun, C. Sturm and F. K. Jondral, “On the single-target accuracy of OFDM radar algorithms,” in {\em IEEE Int. Symp. Person Indoor Mobile Radio Commun. (PIMRC)}, 2011, pp. 794-798.

\bibitem{sturm} C. Sturm and W. Wiesbeck, “Waveform design and signal processing aspects for fusion of wireless communications and radar sensing,” {\em Proc. IEEE}, vol. 99, no. 7, pp. 1236-1259, Jul. 2011.

%%\bibitem{Braun ml} M. Braun, C. Sturm and F. K. Jondral, “Maximum likelihood speed and distance estimation for OFDM radar,” in {\em IEEE Nat. Radar Conf. Proc.}, 2010, pp. 256-261.

\bibitem{no cp} R. F. Tigrek, W. J. A. de Heij and P. van Genderen, “Multi-carrier radar waveform schemes for range and Doppler processing,” in {\em IEEE Nat. Radar Conf. Proc.}, 2009, pp. 1-5.

\bibitem{repeated symbols} G. Hakobyan, M. Girma, X. Li, N. Tammireddy and B. Yang, “Repeated symbols OFDM-MIMO radar at 24 GHz,” in {\em Euro. Radar Conf. (EuRAD)}, 2016, pp. 249-252.

\bibitem{fd and delta_f} G. Hakobyan and B. Yang, “A novel intercarrier-interference free signal processing scheme for OFDM radar,” {\em IEEE Trans. Veh. Technol.}, vol. 67, no. 6, pp. 5158-5167, Jun. 2018.

\bibitem{self interference} A. Tang, S. Li and X. Wang, “Self-interference-resistant IEEE 802.11ad-based joint communication and automotive radar design,” {\em IEEE J. Sel. Top. Sign. Proces.}, vol. 15, no. 6, pp. 1484-1499, Nov. 2021.

\bibitem{vcp} K. Wu, J. A. Zhang, X. Huang and Y. J. Guo, “Integrating low-complexity and flexible sensing into communication systems,” {\em IEEE J. Sel. Areas Commun.}, vol. 40, no. 6, pp. 1873-1889, Jun. 2022.

\bibitem{3GPP 38.211} 3GPP, TS 38.211 (V17.3.0), “Physical channels and modulation,” Sep. 2022.

\bibitem{3GPP 38.901} 3GPP, TR 38.901 (V17.0.0), “Study on channel model for frequencies from 0.5 to 100 GHz,” Mar. 2022.
\end{thebibliography}
\end{document}